\documentclass[12pt]{article}

\usepackage{graphicx}
\usepackage{dcolumn}
\usepackage{bm}
\usepackage{amsmath}
\usepackage{cite}

\newcommand{\be}{\begin{eqnarray}}
\newcommand{\ee}{\end{eqnarray}}
\newcommand{\ba}{\begin{array}}
\newcommand{\ea}{\end{array}}
\catcode`\@=11
\def\lsim{\mathrel{\mathpalette\@versim<}}
\def\gsim{\mathrel{\mathpalette\@versim>}}
\def\@versim#1#2{\vcenter{\offinterlineskip
\ialign{$\m@th#1\hfil##\hfil$\crcr#2\crcr\sim\crcr } }}
\catcode`\@=12

\begin{document}
\begin{center}
{\Large\bf Lepton flavour violating processes in an $S_3$-symmetric model.}
\end{center} 
\vspace{0.7cm}
\begin{center} 
{\sc A. Mondrag\'on} ,
{\sc M. Mondrag\'on }, and {\sc E. Peinado }
\vspace{0.4cm}

Instituto de    F\'{\i}sica, Universidad Nacional Aut\'onoma
  de M\'exico,
    Apdo. Postal 20-364,  01000 M{\'e}xico D.F., \ M{\'e}xico.
\end{center}
\noindent
A variety of lepton flavour violating effects related to the recent
discovery of neutrino oscillations and mixings is here systematically
discussed in terms of an $S_{3}$-flavour permutational symmetry. After
presenting some relevant results on lepton masses and mixings,
previously derived in a minimal $S_{3}$-invariant extension of the
Standard Model, we compute the branching ratios of some selected
flavour-changing neutral current processes (FCNC) as well as the
contribution of the exchange of neutral flavour-changing scalar to the
anomaly of the magnetic moment of the muon. We found that the minimal
$S_{3}$-invariant extension of the Standard Model describes
successfully masses and mixings, as well as, all flavour changing
neutral current processes in excellent agreement with experiment.


\vspace{1cm}
\section{Introduction}
In the past nine years, the experiments and observations of flavour
oscillations of solar~\cite{altmann,smy,ahmad,aharmim}, atmospheric~\cite{fukuda,Ashie:2005ik} and reactor~\cite{bemporad,Araki:2004mb} neutrinos,
established beyond reasonable doubt that neutrinos have non-vanishing
masses and mix among themselves much like the quarks do. This is also
the first experimental evidence that lepton flavour is not conserved
in Nature~\cite{jung,Mohapatra:2006gs}.

On the theoretical side, the discovery of neutrino masses and mixings
is the first conclusive evidence of the incompleteness of the Standard
Model, expected on theoretical grounds since long ago~\cite{GellMann:1980vs,yanagida,Schechter:1980gr,Mohapatra:1980yp,Giunti:2007ry,Strumia:2006db,Langacker:2005pfa}. Hence,
the need of extending the Standard Model in a logically consistent and
physically coherent way to allow for lepton flavour violation and a
unified and systematic treatment of the observed hierarchies of masses
and mixings of all fermions. At the same time, it would be highly
desirable to reduce drastically the number of free parameters in the
theory. A flavour symmetry that generates the observed pattern of
fermion mixings and masses could easily satisfy these two apparently
contradictory requirements.

In a recent paper~\cite{kubo1}, we proposed a minimal extension of the
Standard Model in which the permutational symmetry, $S_3$, is assumed
to be an exact flavour symmetry at the weak scale. In a serie of
subsequent papers~\cite{Felix:2006pn,Mondragon:2007af,Mondragon:2007nk,Mondragon:2007jx}, we made a detailed analysis of masses and mixings
in the leptonic sector of the $S_3-$invariant extended model and also
analyzed the flavour changing neutral current (FCNC) processes in the
leptonic sector of the theory~\cite{Mondragon:2007af} as well as the contribution of
the exchange of flavour changing scalars to the anomaly of the
magnetic moment of the muon~\cite{Mondragon:2007nk}.

In this paper, we review, extend and update the results of our previous
analysis of lepton masses and mixings in section 2, flavour changing
neutral currents are discussed in section 3 and, in section 4, the
contribution of the exchange of neutral scalars to the anomaly of the
magnetic moment of the muon is explained in some detail. We end our
paper with a short summary of results and some conclusions.

\section{A Minimal $S_{3}$-invariant Extension of the Standard
  Model}
In the Standard Model, the Higgs and Yukawa sectors which are
responsible for the generation of masses of quarks and charged leptons
do not give mass to the neutrinos. Furthermore, the Yukawa sector of
the Standard Model already has too many parameters whose values can
only be determined from experiment.  These two facts
point to the necessity and convenience of extending the Standard Model
in order to make a unified and
systematic treatment of the observed hierarchies of
masses and mixings of all fermions, as well as the presence or absence of CP violating
phases in the mixing matrices. At the same time, we would also like to
reduce drastically the number of free parameters in the theory. These
two seemingly contradictory demands can be met by means of a flavour
symmetry under which the families transform in a non-trivial fashion.

Recently, we introduced a minimal $S_{3}$-invariant extension of the Standard
  Model~\cite{kubo1} in which we argued that such a flavour
symmetry unbroken at the Fermi scale, is the permutational symmetry of
three objects $S_{3}$. In this model, we imposed $S_{3}$ as a
fundamental symmetry in the matter sector. This assumption led us
necessarily to extend the concept of flavour and generations to the
Higgs sector. Hence, going to the irreducible representations of
$S_{3}$, we added to the Higgs $SU(2)_{L}$ doublet in the
$S_{3}$-singlet representation two more Higgs $SU(2)_{L}$
doublets, which can only belong to the two components of the
$S_{3}$-doublet representation. In this way, all the matter fields in
the minimal $S_{3}$-invariant extension of the Standard Model - Higgs,
quark and lepton fields, including the right handed neutrino fields-
belong to the three dimensional representation ${\bf 1}\oplus{\bf 2}$
of the permutational group $S_{3}$. The quark, lepton and Higgs fields are
\be
\begin{array}{lr}
Q^T=(u_L,d_L)~,~ u_R~,~d_R~,~\\L^T=(\nu_L,e_L)~,~e_R~,~ 
\nu_R~\mbox{ and }~H,
\end{array}
\ee
in an obvious notation. All of these fields have three species, and,
as explained above, we assume that each one form a reducible
representation ${\bf 1}_S\oplus{\bf 2}$. The doublets carry capital indices $I$ and $J$, which run from $1$ to $2$,
and the singlets are denoted by
$Q_3,~u_{3R},~d_{3R},~L_3,~e_{3R},~\nu_{3R}$ and $~H_S$. Note that the subscript $3$ denotes the
singlet representation and not the third generation.

Due to the presence of three Higgs field, the Higgs potential
$V_H(H_S,H_D)$ is more complicate than that of the Standard Model
\be
V_H(H_S,H_D)=V_1+V_2
\ee
where
\be
\label{v1}
\begin{split}
V_1=\mu^2\left[(\bar{H}_{D1}H_{D1})+(\bar{H}_{D2}H_{D2})+(\bar{H}_{3}H_{3})\right]+\\
\frac{1}{2}\lambda_1 \left[(\bar{H}_{D1}H_{D1})+(\bar{H}_{D2}H_{D2})+(\bar{H}_{3}H_{3}) \right]^2
\end{split}
\ee
and
\be
\label{v2}
V_2=\eta_1 (\bar{H}_{3}H_{3})\left[(\bar{H}_{D1}H_{D1})+(\bar{H}_{D2}H_{D2})\right].
\ee

A Higgs potential invariant under $S_{3}$ was first proposed by
Pakvasa and Sugawara~\cite{pakvasa1}, who assumed an additional
reflection symmetry $R:~H_{s}\rightarrow ~-H_{s}$. These authors found
that in addition to the $S_{3}$ symmetry, their Higgs potential has an
accidental symmetry $S_{2}^{\prime}$:$H_{1}\leftrightarrow H_{2}$. The accidental $S_{2}^{\prime}$ symmetry
is also present in our $V_H(H_S,H_D)$, therefore, $\langle H_{1}
\rangle = \langle H_{2} \rangle$.

The most general renormalizable Yukawa interactions of this model are given by
\be
{\cal L}_Y = {\cal L}_{Y_D}+{\cal L}_{Y_U}
+{\cal L}_{Y_E}+{\cal L}_{Y_\nu},
\ee
where
\be
\begin{array}{lll}
{\cal L}_{Y_D} &=&
- Y_1^d \overline{ Q}_I H_S d_{IR} - Y_3^d \overline{ Q}_3 H_S d_{3R}  \\
&  &   -Y^{d}_{2}[~ \overline{ Q}_{I} \kappa_{IJ} H_1  d_{JR}
+\overline{ Q}_{I} \eta_{IJ} H_2  d_{JR}~]\\
&  & -Y^d_{4} \overline{ Q}_3 H_I  d_{IR} - Y^d_{5} \overline{ Q}_I H_I d_{3R} 
+~\mbox{h.c.} ,
\label{lagd}
\end{array}
\ee
\be
\begin{array}{lll}
{\cal L}_{Y_U} &=&
-Y^u_1 \overline{ Q}_{I}(i \sigma_2) H_S^* u_{IR} 
-Y^u_3\overline{ Q}_3(i \sigma_2) H_S^* u_{3R} \\
&  &   -Y^{u}_{2}[~ \overline{ Q}_{I} \kappa_{IJ} (i \sigma_2)H_1^*  u_{JR}
+  \overline{ Q}_{I} \eta_{IJ}(i \sigma_2) H_2^*  u_{JR}~]\\
&  &
-Y^u_{4} \overline{ Q}_{3} (i \sigma_2)H_I^* u_{IR} 
-Y^u_{5}\overline{ Q}_I (i \sigma_2)H_I^*  u_{3R} +~\mbox{h.c.},
\label{lagu}
\end{array}
\ee
\be
\begin{array}{lll}
{\cal L}_{Y_E} &=& -Y^e_1\overline{ L}_I H_S e_{IR} 
-Y^e_3 \overline{ L}_3 H_S e_{3R} \\
&  &  - Y^{e}_{2}[~ \overline{ L}_{I}\kappa_{IJ}H_1  e_{JR}
+\overline{ L}_{I} \eta_{IJ} H_2  e_{JR}~]\\
 &  & -Y^e_{4}\overline{ L}_3 H_I e_{IR} 
-Y^e_{5} \overline{ L}_I H_I e_{3R} +~\mbox{h.c.},
\end{array}
\label{lage}
\ee
\be
\begin{array}{lcl}
{\cal L}_{Y_\nu} &=& -Y^{\nu}_1\overline{ L}_I (i \sigma_2)H_S^* \nu_{IR} 
-Y^\nu_3 \overline{ L}_3(i \sigma_2) H_S^* \nu_{3R} \\
&  &   -Y^{\nu}_{2}[~\overline{ L}_{I}\kappa_{IJ}(i \sigma_2)H_1^*  \nu_{JR}
+ \overline{ L}_{I} \eta_{IJ}(i \sigma_2) H_2^*  \nu_{JR}~]\\
 &  & -Y^\nu_{4}\overline{ L}_3(i \sigma_2) H_I^* \nu_{IR} 
-Y^\nu_{5} \overline{ L}_I (i \sigma_2)H_I^* \nu_{3R}+~\mbox{h.c.},
\label{lagnu}
\end{array}
\ee
and
\be
\kappa = \left( \begin{array}{cc}
0& 1\\ 1 & 0\\
\end{array}\right)~~\mbox{and}~~
\eta = \left( \begin{array}{cc}
1& 0\\ 0 & -1\\
\end{array}\right).
\label{kappa}
\ee
Furthermore, we add to the Lagrangian the Majorana mass terms for
the right-handed neutrinos \be {\cal L}_{M} = -M_1 \nu_{IR}^T C
\nu_{IR} -M_3 \nu_{3R}^T C \nu_{3R}.
\label{majo}
\ee
With these assumptions, the Yukawa interactions, eqs. (\ref{lagd})-(\ref{lagnu}) yield mass matrices,
for all fermions in the theory, of the general form~\cite{kubo1}
\be
{\bf M} = \left( \begin{array}{ccc}
\mu_{1}+\mu_{2} & \mu_{2} & \mu_{5} 
\\  \mu_{2} & \mu_{1}-\mu_{2} &\mu_{5}
  \\ \mu_{4} & \mu_{4}&  \mu_{3}
\end{array}\right).
\label{general-m}
\ee
The Majorana mass for the left handed neutrinos $\nu_{L}$ is generated
from (\ref{majo}) by the see-saw mechanism,
\be
{\bf M_{\nu}} = {\bf M_{\nu_D}}\tilde{{\bf M}}^{-1}({\bf M_{\nu_D}})^T,
\label{seesaw}
\ee
where $\tilde{{\bf M}}=\mbox{diag}(M_1,M_1,M_3)$.
In principle, all entries in the mass matrices can be complex since
there is no restriction coming from the flavour symmetry $S_{3}$.
The mass matrices are diagonalized by bi-unitary transformations as
\be
\begin{array}{rcl}
U_{d(u,e)L}^{\dag}{\bf M}_{d(u,e)}U_{d(u,e)R} 
&=&\mbox{diag} (m_{d(u,e)}, m_{s(c,\mu)},m_{b(t,\tau)}),
\\ 
\\
U_{\nu}^{T}{\bf M_\nu}U_{\nu} &=&
\mbox{diag} (m_{\nu_1},m_{\nu_2},m_{\nu_3}).
\end{array}
\label{unu}
\ee
The entries in the diagonal matrices may be complex, so the physical
masses are their absolute values.

The mixing matrices are, by definition,
\be
\begin{array}{ll}
V_{CKM} = U_{uL}^{\dag} U_{dL},& V_{PMNS} = U_{eL}^{\dag} U_{\nu} K.
\label{ckm1}
\end{array}
\ee
where $K$ is the diagonal matrix of the Majorana phase factors.
\section{The mass matrices in the leptonic sector and $Z_{2}$
  symmetry}
The number of free parameters in the leptonic sector may be further
reduced by means of an Abelian $Z_2$ symmetry. A possible set of
charge assignments of $Z_{2}$, compatible with the experimental
data on masses and mixings in the leptonic sector is given in Table~\ref{table1}.
\begin{table}
\caption{\label{table1}$Z_2$ assignment in the leptonic sector.}
\begin{center}
\begin{tabular}{|c|c|}
\hline
 $-$ &  $+$
\\ \hline

$H_S, ~\nu_{3R}$ & $H_I, ~L_3, ~L_I, ~e_{3R},~ e_{IR},~\nu_{IR}$
\\ \hline
\end{tabular}
\end{center}
\end{table}
These $Z_2$ assignments forbid the following Yukawa couplings
\be
 Y^e_{1} = Y^e_{3}= Y^{\nu}_{1}= Y^{\nu}_{5}=0.
\label{zeros}
\ee
Hence, the corresponding entries in the mass matrices vanish, {\it
  i.e.},
\[\mu_{1}^{e}=\mu_{3}^{e}=0,\]
and
\[
\mu_{1}^{\nu}=\mu_{5}^{\nu}=0.
\]

\bigskip

\begin{center}
{\it The mass matrix of the neutrinos}
\end{center}
According to the $Z_2$ selection rule, Eq (\ref{zeros}), the mass
matrix of the Dirac neutrinos takes the form
\be
{\bf M_{\nu_D}} = \left( \begin{array}{ccc}
\mu^{\nu}_{2} & \mu^{\nu}_{2} & 0
\\  \mu^{\nu}_{2} & -\mu^{\nu}_{2} &0
  \\ \mu^{\nu}_{4} & \mu^{\nu}_{4}&  \mu^{\nu}_{3}
\end{array}\right).
\label{neutrinod-m}
\ee
Then, the mass matrix for the left-handed Majorana neutrinos, ${\bf
  M_{\nu}}$, obtained
from the see-saw mechanism, ${\bf M_{\nu}} = {\bf M_{\nu_D}}\tilde{{\bf M}}^{-1} 
({\bf M_{\nu_D}})^T$, is
\be
{\bf M_{\nu}} = 
\left( \begin{array}{ccc}
2 (\rho^{\nu}_{2})^2 & 0 & 
2 \rho^{\nu}_2 \rho^{\nu}_{4}
\\ 0 & 2 (\rho^{\nu}_{2})^2 & 0
  \\ 2 \rho^{\nu}_2 \rho^{\nu}_{4} & 0  &  
2 (\rho^{\nu}_{4})^2 +
(\rho^{\nu}_3)^2
\end{array}\right),
\label{m-nu}
\ee
where $\rho_2^\nu =(\mu^{\nu}_2)/M_{1}^{1/2}$,  $\rho_4^\nu
=(\mu^{\nu}_4)/M_{1}^{1/2}$ and $\rho_3^\nu
=(\mu^{\nu}_3)/M_{3}^{1/2}$; $M_{1}$ and $M_{3}$ are the masses of
the right handed neutrinos appearing in (\ref{majo}).

The non-Hermitian, complex, symmetric neutrino mass matrix $M_{\nu}$ may be brought
to a diagonal form by a unitary transformation, as
\be
U_{\nu}^{T}M_{\nu}U_{\nu}=\mbox{diag}\left(|m_{\nu_{1}}|e^{i\phi_{1}},|m_{\nu_{2}}|e^{i\phi_{2}},|m_{\nu_{3}}|e^{i\phi_{\nu}}\right),
\label{diagneutrino}
\ee
where $U_{\nu}$ is the matrix that diagonalizes the matrix
$M_{\nu}^{\dagger}M_{\nu}$.

Written in polar form, the matrix $U_{\nu}$ takes the form
\be
U_{\nu}=\left(\ba{ccc} 
1& 0 & 0 \\
0 & 1 & 0 \\
0 & 0 & e^{i\delta_{\nu}} 
\ea\right)\left(
\begin{array}{ccc}
\cos \eta & \sin \eta & 0 \\
0 & 0 & 1 \\
-\sin \eta  & \cos \eta &0
\end{array}
\right),
\label{ununew0}
\ee
if we require that the defining Eq. (\ref{diagneutrino}) be satisfied as an
identity, we may solve the resulting equations for $\sin \eta$ and
$\cos \eta$ in terms of the neutrino masses. This allows us to
reparametrize the matrices $M_\nu$ and $U_\nu$ in terms of the complex
neutrino masses,
\be
M_{\nu} = 
\left( \begin{array}{ccc}
m_{\nu_{3}} & 0 & \sqrt{(m_{\nu_{3}}-m_{\nu_{1}})(m_{\nu_{2}}-m_{\nu_{3}})}e^{-i\delta_{\nu}}
\\ 
0 &m_{\nu_{3}}  & 0
\\
\sqrt{(m_{\nu_{3}}-m_{\nu_{1}})(m_{\nu_{2}}-m_{\nu_{3}})} e^{-i\delta_{\nu}} & 0  & (m_{\nu_{1}}+m_{\nu_{2}}-m_{\nu_{3}})e^{-2i\delta_{\nu}}
\end{array}\right)
\label{m-nu2}
\ee
and 
\be
U_{\nu}=\left(\ba{ccc} 
1& 0 & 0 \\
0 & 1 & 0 \\
0 & 0 & e^{i\delta_{\nu}} 
\ea\right)\left(
\begin{array}{ccc}
\sqrt{\frac{m_{\nu_{2}}-m_{\nu_{3}}}{m_{\nu_{2}}-m_{\nu_{1}}}} & \sqrt{\frac{m_{\nu_{3}}-m_{\nu_{1}}}{m_{\nu_{2}}-m_{\nu_{1}}}} & 0 \\
0 & 0 & 1 \\
-\sqrt{\frac{m_{\nu_{3}}-m_{\nu_{1}}}{m_{\nu_{2}}-m_{\nu_{1}}}}  & \sqrt{\frac{m_{\nu_{2}}-m_{\nu_{3}}}{m_{\nu_{2}}-m_{\nu_{1}}}} &0
\end{array}
\right),
\label{ununew}
\ee
The unitarity of $U_{\nu}$ constrains $\sin \eta$ to be real and thus 
$|\sin \eta|\leq 1$, this condition fixes the phases $\phi_{1}$ and
$\phi_{2}$ as
\be
|m_{\nu_{1}}|\sin \phi_{1}=|m_{\nu_{2}}|\sin \phi_{2}=|m_{\nu_{3}}|\sin \phi_{\nu}.
\label{phase-condition}
\ee
The only free parameters in the matrices $M_\nu$ and $U_\nu$, are the  phase $\phi_{\nu}$, implicit in $m_{\nu_{1}}$,
$m_{\nu_{2}}$ and $m_{\nu_{3}}$, and the Dirac phase $\delta_{\nu}$.
\begin{center}{\it The mass matrix of the charged leptons}\end{center}
The mass matrix of the charged leptons takes the form
\be
M_{e} = m_{\tau}\left( \begin{array}{ccc}
\tilde{\mu}_{2} & \tilde{\mu}_{2} & \tilde{\mu}_{5} 
\\  \tilde{\mu}_{2} &-\tilde{\mu}_{2} &\tilde{\mu}_{5}
  \\ \tilde{\mu}_{4} & \tilde{\mu}_{4}& 0
\end{array}\right).
\label{charged-leptons-m}
\ee
The unitary matrix $U_{eL}$ that enters in the definition of the
mixing matrix, $V_{PMNS}$, is calculated from
\be
U_{eL}^{\dag}M_{e}M_{e}^{\dag}U_{eL}=\mbox{diag}(m_{e}^{2},m_{\mu}^{2},m_{\tau}^{2}),
\ee
where $m_{e}$, $m_{\mu}$ and $m_{\tau}$ are the masses of the charged
leptons.

The parameters $|\tilde{\mu}_{2}|$, $|\tilde{\mu}_{4}|$ and
$|\tilde{\mu}_{5}|$ may readily be expressed in terms of the charged
lepton masses. The resulting expression for $M_e$, written to order
$\left(m_{\mu}m_{e}/m_{\tau}^{2}\right)^{2}$ and
$x^{4}=\left(m_{e}/m_{\mu}\right)^4$ is
\bigskip
\be
M_{e}\approx m_{\tau} \left( 
\begin{array}{ccc}
\frac{1}{\sqrt{2}}\frac{\tilde{m}_{\mu}}{\sqrt{1+x^2}} & \frac{1}{\sqrt{2}}\frac{\tilde{m}_{\mu}}{\sqrt{1+x^2}} & \frac{1}{\sqrt{2}} \sqrt{\frac{1+x^2-\tilde{m}_{\mu}^2}{1+x^2}} \\ \\
 \frac{1}{\sqrt{2}}\frac{\tilde{m}_{\mu}}{\sqrt{1+x^2}} &-\frac{1}{\sqrt{2}}\frac{\tilde{m}_{\mu}}{\sqrt{1+x^2}}  & \frac{1}{\sqrt{2}} \sqrt{\frac{1+x^2-\tilde{m}_{\mu}^2}{1+x^2}} \\ \\
\frac{\tilde{m}_{e}(1+x^2)}{\sqrt{1+x^2-\tilde{m}_{\mu}^2}}e^{i\delta_{e}} & \frac{\tilde{m}_{e}(1+x^2)}{\sqrt{1+x^2-\tilde{m}_{\mu}^2}}e^{i\delta_{e}} & 0
\end{array}
\right).
\label{emass}
\ee
This approximation is numerically exact up to order $10^{-9}$ in units
of the $\tau$ mass. Notice that this matrix has no free parameters
other than the Dirac phase $\delta_e$.

The unitary matrix $U_{eL}$ that diagonalizes $M_{e}M_{e}^{\dagger}$ and
enters in the definition of the neutrino mixing matrix $V_{PMNS}$ may
be written in polar form as
\be
\ba{l}
U_{eL}= \left(\ba{ccc} 
1& 0 & 0 \\
0 & 1 & 0 \\
0 & 0 & e^{i\delta_{e}}
\ea\right) \left(
\ba{ccc}
O_{11}& -O_{12}& O_{13} \\
-O_{21}& O_{22}& O_{23} \\
-O_{31}& -O_{32}& O_{33} 
\ea
\right)~,
\ea
\label{unitary-leptons}
\ee
where the orthogonal matrix ${\bf O}_{eL}$ in the right-hand side  of
eq. (\ref{unitary-leptons}), written to the same order of magnitude as
$M_e$, is
\be
{\bf O}_{eL}\approx
\left(
\ba{ccc}
\frac{1}{\sqrt{2}}x
\frac{(
1+2\tilde{m}_{\mu}^2+4x^2+\tilde{m}_{\mu}^4+2\tilde{m}_{e}^2
)}{\sqrt{1+\tilde{m}_{\mu}^2+5x^2-\tilde{m}_{\mu}^4-\tilde{m}_{\mu}^6+\tilde{m}_{e}^2+12x^4}}&
-\frac{1}{\sqrt{2}}\frac{(1-2\tilde{m}_{\mu}^2+\tilde{m}_{\mu}^4-2\tilde{m}_{e}^2)}{\sqrt{1-4\tilde{m}_{\mu}^2+x^2+6\tilde{m}_{\mu}^4-4\tilde{m}_{\mu}^6-5\tilde{m}_{e}^2}}
& \frac{1}{\sqrt{2}} \\ \\
-\frac{1}{\sqrt{2}}x
\frac{(
1+4x^2-\tilde{m}_{\mu}^4-2\tilde{m}_{e}^2
)}{\sqrt{1+\tilde{m}_{\mu}^2+5x^2-\tilde{m}_{\mu}^4-\tilde{m}_{\mu}^6+\tilde{m}_{e}^2+12x^4}}
&
\frac{1}{\sqrt{2}}\frac{(1-2\tilde{m}_{\mu}^2+\tilde{m}_{\mu}^4)}{\sqrt{1-4\tilde{m}_{\mu}^2+x^2+6\tilde{m}_{\mu}^4-4\tilde{m}_{\mu}^6-5\tilde{m}_{e}^2}}
& \frac{1}{\sqrt{2}} \\ \\
-\frac{\sqrt{1+2x^2-\tilde{m}_{\mu}^2-\tilde{m}_{e}^2}(1+\tilde{m}_{\mu}^2+x^2-2\tilde{m}_{e}^2)}{\sqrt{1+\tilde{m}_{\mu}^2+5x^2-\tilde{m}_{\mu}^4-\tilde{m}_{\mu}^6+\tilde{m}_{e}^2+12x^4}} & -x\frac{(1+x^2-\tilde{m}_{\mu}^2-2\tilde{m}_{e}^2)\sqrt{1+2x^2-\tilde{m}_{\mu}^2-\tilde{m}_{e}^2}}{\sqrt{1-4\tilde{m}_{\mu}^2+x^2+6\tilde{m}_{\mu}^4-4\tilde{m}_{\mu}^6-5\tilde{m}_{e}^2}} &\frac{\sqrt{1+x^2}\tilde{m}_{e}\tilde{m}_{\mu}}{\sqrt{1+x^2-\tilde{m}_{\mu}^2}}
\ea
\right)~,
\label{unitary-leptons-2}
\ee
where, as before, $\tilde{m_{\mu}}=m_{\mu}/m_{\tau}$,
$\tilde{m_{e}}=m_{e}/m_{\tau}$ and $x=m_{e}/m_{\mu}$.

\bigskip

\begin{center}
{\it The neutrino mixing matrix}
\end{center}

The neutrino mixing matrix $V_{PMNS}$, is the product
$U_{eL}^{\dagger}U_{\nu}K$, where $K$ is the diagonal matrix of the
Majorana phase factors, defined by
\be
diag(m_{\nu_{1}},m_{\nu_{2}},m_{\nu_{3}})=K^{\dagger}diag(|m_{\nu_{1}}|,|m_{\nu_{2}}|,|m_{\nu_{3}}|)K^{\dagger}.
\ee
Except for an overall phase factor $e^{i\phi_{1}}$, which can be
ignored, $K$ is 
\be
K=diag(1,e^{i\alpha},e^{i\beta}),
\ee
where $\alpha=1/2(\phi_{1}-\phi_{2})$ and
$\beta=1/2(\phi_{1}-\phi_{\nu})$ are the Majorana phases. 

Therefore, the theoretical mixing matrix $V_{PMNS}^{th}$, is given by
\be
V_{PMNS}^{th}=
\left(
\ba{ccc}
O_{11}\cos \eta + O_{31}\sin \eta e^{i\delta} & O_{11}\sin
\eta-O_{31} \cos \eta e^{i\delta} & -O_{21}  \\ \\
-O_{12}\cos \eta + O_{32}\sin \eta e^{i\delta} & -O_{12}\sin
\eta-O_{32}\cos \eta e^{i\delta} & O_{22} \\ \\
O_{13}\cos \eta - O_{33}\sin \eta e^{i\delta} & O_{13}\sin
\eta+O_{33}\cos \eta e^{i\delta} & O_{23} 
\ea
\right)
 \times K,
\label{vpmns2}
\ee
where $\cos \eta$ and $\sin \eta$ are given in eqs  and
(\ref{ununew0}) and (\ref{ununew}) ,
$O_{ij}$ are given in eq (\ref{unitary-leptons}) and
(\ref{unitary-leptons-2}), and $\delta=\delta_{\nu}-\delta_{e}$. 

To find the relation of our results
with the neutrino mixing angles we make use of the equality of the
absolute values of the elements of $V_{PMNS}^{th}$ and
$V_{PMNS}^{PDG}$~\cite{PDG}, that is
\be
|V_{PMNS}^{th}|=|V_{PMNS}^{PDG}|.
\ee
This relation allows us to derive expressions for the mixing angles
in terms of the charged lepton and neutrino masses.

The magnitudes of the reactor and atmospheric mixing angles,
$\theta_{13}$ and $\theta_{23}$, are determined by the masses of the
charged leptons only. Keeping only terms of order $(m_{e}^2/m_{\mu}^2)$ and
$(m_{\mu}/m_{\tau})^4$, we get
\be
\ba{lr}
\sin \theta_{13}\approx \frac{1}{\sqrt{2}}x
\frac{(
1+4x^2-\tilde{m}_{\mu}^4)}{\sqrt{1+\tilde{m}_{\mu}^2+5x^2-\tilde{m}_{\mu}^4}}
, &
\sin \theta_{23}\approx  \frac{1}{\sqrt{2}}\frac{1+\frac{1}{4}x^2-2\tilde{m}_{\mu}^2+\tilde{m}_{\mu}^4}{\sqrt{1-4\tilde{m}_{\mu}^2+x^2+6\tilde{m}_{\mu}^4}}.
\ea
\ee
The magnitude of the solar angle depends on charged lepton and
neutrino masses, as well as, the Dirac and Majorana phases
\be
 |\tan \theta_{12}|^2= \frac{\displaystyle{m_{\nu_{2}}-m_{\nu_{3}}}}{
\displaystyle{m_{\nu_{3}}-m_{\nu_{1}}}}\left(\frac{1-2\frac{O_{11}}{O_{31}}\cos \delta\sqrt{\frac{\displaystyle{m_{\nu_{3}}-m_{\nu_{1}}}}{
\displaystyle{m_{\nu_{2}}-m_{\nu_{3}}}}}+\left(\frac{O_{11}}{O_{31}}\right)^2\frac{\displaystyle{m_{\nu_{3}}-m_{\nu_{1}}}}{
\displaystyle{m_{\nu_{2}}-m_{\nu_{3}}}}}{1+2\frac{O_{11}}{O_{31}}\cos \delta\sqrt{\frac{\displaystyle{m_{\nu_{2}}-m_{\nu_{3}}}}{
\displaystyle{m_{\nu_{3}}-m_{\nu_{1}}}}}+\left(\frac{O_{11}}{O_{31}}\right)^2\frac{\displaystyle{m_{\nu_{2}}-m_{\nu_{3}}}}{
\displaystyle{m_{\nu_{3}}-m_{\nu_{1}}}}
}\right)
.
\label{tan2}
\ee
The dependence of $\tan \theta_{12}$ on the Dirac phase $\delta$, see
(\ref{tan2}), is very weak, since $O_{31}\sim 1$ but $O_{11}\sim
1/\sqrt{2}(m_e/m_\mu)$. Hence, we may neglect it when comparing
(\ref{tan2}) with the data on neutrino mixings.

The dependence of $\tan \theta_{12}$ on the phase $\phi_{\nu}$ and the
physical masses of the neutrinos enters through the ratio of the
neutrino mass differences, it can be made explicit with the help of
the unitarity constraint on $U_{\nu}$, 
eq. (\ref{phase-condition}),
\be
\frac{\displaystyle{m_{\nu_{2}}-m_{\nu_{3}}}}{
\displaystyle{m_{\nu_{3}}-m_{\nu_{1}}}}=
\frac{(|m_{\nu_{2}}|^2-|m_{\nu_{3}}|^{2}\sin^{2}\phi_{\nu})^{1/2}-|m_{\nu_{3}}|
  |\cos
    \phi_{\nu}|}
{(|m_{\nu_{1}}|^{2}-|m_{\nu_{3}}|^{2}\sin^{2}\phi_{\nu})^{1/2}+|m_{\nu_{3}}|
  |\cos
    \phi_{\nu}|}.
\label{tansq}
\ee
Similarly, the Majorana phases are given by
\be
\ba{l}
\sin 2\alpha=\sin(\phi_{1}-\phi_{2})=
\frac{|m_{\nu_{3}}|\sin\phi_{\nu}}{|m_{\nu_{1}}||m_{\nu_{2}}|}\times
\\
\left(\sqrt{|m_{\nu_{2}}|^2-|m_{\nu_{3}}|^{2}\sin^{2}\phi_{\nu}}+\sqrt{|m_{\nu_{1}}|^{2}-|m_{\nu_{3}}|^{2}\sin^{2}\phi_{\nu}}\right),
\ea
\ee
\be
\ba{l}
\sin 2\beta=\sin(\phi_{1}-\phi_{\nu})=
\\
 \frac{\sin\phi_{\nu}}{|m_{\nu_{1}}|}\left(|m_{\nu_{3}}|\sqrt{1-\sin^{2}\phi_{\nu}}+\sqrt{|m_{\nu_{1}}|^{2}-|m_{\nu_{3}}|^{2}\sin^{2}\phi_{\nu}}\right).
\ea
\ee
A more complete and detailed discussion of the Majorana phases in the
neutrino mixing matrix $V_{PMNS}$ obtained in our model is given by 
J. Kubo~\cite{kubo-u}.

\section{Neutrino masses and mixings}
The determination of neutrino oscillation parameters~\cite{Maltoni:2004ei,schwetz,Gonzalez-Garcia:2007ib,chooz,eitel} has finally
entered the high precision age, with many experiments underway and a
new generation coming. The results of two global analysis of neutrino
oscillations, updated to september 2007, are summarized in
Table~\ref{parameters} taken from~\cite{Maltoni:2004ei}. In this section, numerical values of the neutrino masses and
mixing angles will be obtained from the theoretical expressions derived
in the previous section and the numerical values of the neutrino
oscillation parameters given in Table~\ref{parameters}.
\begin{table}[t] \centering
    
\begin{tabular}{|c|c|c|c|}
        \hline
        parameter & best fit & 2$\sigma$ & 3$\sigma$ 
        \\
        \hline
        $\Delta m^2_{21}  [10^{-5} ~eV ]$
        & 7.6 & 7.3--8.1 & 7.1--8.3 \\
        $\Delta m^2_{31} [10^{-3} ~eV ]$
        & 2.4 & 2.1--2.7 & 2.0--2.8 \\
        $\sin^2\theta_{12}$
        & 0.32 & 0.28--0.37 & 0.26--0.40\\
        $\sin^2\theta_{23}$
        & 0.50 & 0.38--0.63 & 0.34--0.67\\
        $\sin^2\theta_{13}$
        & 0.007 &  $\leq$ 0.033 & $\leq$ 0.050 \\
        \hline
\end{tabular}
\caption{\label{parameters} 
  %
  Best-fit values, 2$\sigma$ and
  3$\sigma$ intervals (1 d.o.f) for the
  three-flavour neutrino oscillation parameters from global data
  including solar, atmospheric, reactor (KamLAND and CHOOZ) and
  accelerator (K2K and MINOS) experiments.} 
\end{table}
In the present $S_3$-invariant extension of the Standard Model, the
experimental restriction $|\Delta m_{12}^2|<|\Delta m_{13}^2|$ implies
an inverted neutrino mass spectrum
$|m_{\nu_{3}}|<|m_{\nu_{1}}|<|m_{\nu_{2}}|$~\cite{kubo1}.

In this model, the reactor and atmospheric mixing angles, $\theta_{13}$
and $\theta_{23}$, are determined by the masses of the charged leptons
only, in very good agreement with the experimental values
\be
\begin{array}{ll}
(\sin^{2}\theta_{13})^{th}=1.1\times 10^{-5}, &(\sin^2
  \theta_{13})^{exp} \leq 0.028 
\end{array}
\label{cond1}
\ee
and 
\be
\begin{array}{ll}
(\sin^{2}\theta_{23})^{th}=0.5, &(\sin^2
  \theta_{23})^{exp}=0.5^{+0.06}_{-0.05}.
\end{array}
\label{cond2}
\ee
As can be seen from equations (~\ref{tan2}) and (~\ref{tansq}), the solar angle
is sensitive to the differences of the squared neutrino masses and the
phase $\phi_{\nu}$ but is only weakly sensitive to the charged lepton
masses. If the small terms proportional to $O_{11}$ and $O_{11}^2$ are
neglected in (~\ref{tan2}), we obtain
\be
\tan^2 \theta_{12} =
\frac{(\Delta m_{12}^2+\Delta m_{13}^2+|m_{\nu_{3}}|^{2}\cos^2 \phi_\nu)^{1/2}-|m_{\nu_{3}}|\cos \phi_\nu
  }
{(\Delta m_{13}^2+|m_{\nu_{3}|}^{2}\cos^2 \phi_\nu)^{1/2}+|m_{\nu_{3}}|\cos \phi_\nu
  }
\ee
From this equation, we may readily derive expressions for the
neutrino masses in terms of $\tan \theta_{12}$, $\cos \phi_\nu$ and the
differences of the squared neutrino masses
\be
|m_{\nu_{3}}|=\frac{\sqrt{\Delta m_{13}^2}}{2 \tan
  \theta_{12}\cos \phi_\nu}\frac{1-\tan^4 \theta_{12}+r^2}{\sqrt{1+\tan^2 \theta_{12}} \sqrt{1+\tan^2 \theta_{12}+r^2}}
\ee
and 
\bigskip
\be
\ba{l}
|m_{\nu_{1}}|=\sqrt{|m_{\nu_{3}}|^2+\Delta m_{13}^2}, \\
\\
|m_{\nu_{2}}|=\sqrt{|m_{\nu_{3}}|^2+\Delta m_{13}^2(1+r^2)}
\ea
\ee
where $r^2=\Delta m_{12}^2/\Delta m_{13}^2\approx 3\times 10^{-2}$. 

As $r^2<<1$, the sum of the neutrino masses is
\begin{eqnarray}
\sum_{i=1}^{3} |m_{\nu_{i}}|\approx\frac{\sqrt{\Delta m_{13}^2}}{2\cos \phi_{\nu}
  \tan \theta_{12}}\left(1+2\sqrt{1+2\tan^2 \theta_{12}\left(2 \cos^2 \phi_\nu-1\right)+\tan^4 \theta_{12}}-\tan^2 \theta_{12}\right). \nonumber \\ 
\end{eqnarray}
The most restrictive cosmological upper bound for this sum is~\cite{Seljak}
\be
\sum |m_{\nu}|\leq 0.17 eV.
\label{cosmup}
\ee
This upper bound and the experimentally determined values of
$\tan \theta_{12}$ and $\Delta m_{ij}^{2}$, give a lower bound
for $\cos \phi_{\nu}$,
\be
\cos \phi_{\nu}\geq 0.55
\ee
or $0\leq \phi_\nu \leq 57^{\circ}$.

The neutrino masses $|m_{\nu_i}|$
assume their minimal values when $\cos \phi_\nu=1$. When $\cos
\phi_\nu$ takes values in the range $0.55 \leq \cos \phi \leq 1$, the
neutrino masses change very slowly with $\cos \phi_\nu$. In the
absence of experimental information we will assume that $\phi_\nu$ vanishes.
Hence, setting $\phi_\nu=0$ in our formula, we find
\be
\begin{array}{lcr}
m_{\nu_1}= 0.052~eV & m_{\nu_2}=0.053 ~eV & m_{\nu_3}= 0.019~eV.
\end{array}
\ee
The computed sum of the neutrino masses is
\be
\left(\sum_{i=1}^3 |m_{\nu_{i}}|\right)^{th}=0.13 ~eV,
\ee
below the cosmological upper bound given in eq. (\ref{cosmup}), as
expected, since we used the cosmological bound to fix the bound on
$\cos \phi_{\nu}$.

Now, we may compare our results with other bounds on the neutrino
masses.

\noindent
The effective Majorana mass in neutrinoless double beta decay $\langle
m_{2\beta} \rangle$, is defined as~\cite{elliott}
\be
\langle m_{2\beta} \rangle =|\sum_{i=1}^3 V_{ei}^2 m_{\nu_{i}}|.
\ee
\noindent
The most stringent bound on $\langle m_{2\beta} \rangle$, obtained
from the analysis of the data collected by the Heidelberg-Moscow
experiment on neutrinoless double beta decay in enriched Ge~\cite{KlapdorKleingrothaus:2000sn},
is
\be
\langle m_{2\beta} \rangle < 0.3~eV.
\ee
In our model, and assuming that the Majorana phases vanish, we get
\be
\langle m_{2\beta} \rangle ^{th}=0.053 ~eV
\ee
well below the experimental upper bound.

The most restrictive direct neutrino measurement involving electron
type neutrinos, is based on fitting the shape of the beta spectrum~\cite{eitel}. In
such measurement, the quantity
\be
\bar{m}_{\nu_{e}}=\sqrt{\sum_{i}|V_{ei}|^2m_{\nu_{i}}}
\ee
is determined or constrained. A very restrictive upper bound for this
sum is obtained from nucleosynthesis processes~\cite{Fields:1995mz,Kolb:1991sn}
\be
\left(\bar{m}_{\nu_{e}}\right)^{exp}< 0.37 ~eV.
\label{betadecay}
\ee
\noindent
From eq. (\ref{cond1}) and (\ref{cond2}), we obtain
\be
\left(\bar{m}_{\nu_{e}}\right)^{th}= 0.053 ~eV.
\ee
again, well below the experimental upper bound given in eq. (\ref{betadecay})

\section{Lepton flavour violating processes}

It is well known that models with more than one Higgs $SU(2)$ doublet
may in general, have tree-level flavour changing neutral currents
(FCNC)~\cite{Glashow:1976nt,Paschos:1976ay}. Lepton flavour violating
couplings, due to FCNC, naturally appear in the minimal
$S_{3}$-invariant extension of the Standard Model, since in this
extended model there are three Higgs $SU(2)$ doublets, one per
generation, coupled to all fermions. In a flavour labeled, symmetry
adapted weak basis, the flavour changing Yukawa couplings may be written as
\be
\ba{lcl}
\hspace{-8pt}{\cal L}^{\rm FCNC}_{Y} =
\left(\overline{E}_{aL} Y_{a b}^{ES} E_{bR}+
\overline{U}_{aL} Y_{a b}^{US} U_{bR}
+\overline{D}_{aL} Y_{a b}^{DS} D_{bR}\right)H_S^0 \\ \\
~+\left(\overline{E}_{aL} Y_{a b}^{E1} E_{bR}+
\overline{U}_{aL} Y_{a b}^{U1} U_{bR}
+\overline{D}_{aL} Y_{a b}^{D1} D_{bR}\right)H_1^0+ \\ \\
\left(\overline{E}_{aL} Y_{a b}^{E2} E_{bR}+
\overline{U}_{aL} Y_{a b}^{U2} U_{bR}
+\overline{D}_{aL} Y_{a b}^{D2} D_{bR}\right)H_2^0+\mbox{h.c.}
\ea
\label{fcnf-lept}
\ee
in this expression, the entries in the column matrices $E's$, $U's$ and $D's$ are
the left and right fermion fields and $Y_{ab}^{(E,U,D)S}$,
$Y_{ab}^{(E,U,D)1,2}$ are $3\times 3$ matrices of the Yukawa couplings
of the fermion fields to the neutral Higgs fields $H_{s}^0$ and
$H_{1,2}^0$ in the $S_3$-singlet and doublet representations respectively.

In this basis, the Yukawa couplings of the Higgs fields to each
family of fermions may be written in terms of matrices
${\cal{M}}_{Y}^{(e,u,d)}$, which give rise to the corresponding mass
matrices $M^{(e,u,d)}$ when the gauge symmetry is spontaneously
broken. From this relation we may calculate the flavour changing
Yukawa couplings in terms of the fermion masses and the vacuum
expectation values of the neutral Higgs fields.

The matrix ${\cal{M}}_{Y}^e$ of the charged leptons is written in terms of the
matrices of the Yukawa couplings of the charged leptons as
\be
{\cal{M}}_{Y}^e=Y_{w}^{E1} H^0_{1}+Y_{w}^{E2} H^0_{2},
\ee
where, the index $w$ means that the Yukawa matrices are
defined in the weak basis, 
\be
Y_{w}^{E1}=\frac{m_{\tau} }{v_1}\left(
\begin{array}{ccc}
0 & \frac{1}{\sqrt{2}}\frac{\tilde{m}_{\mu}}{\sqrt{1+x^2}} & \frac{1}{\sqrt{2}} \sqrt{\frac{1+x^2-\tilde{m}_{\mu}^2}{1+x^2}} \\ \\
 \frac{1}{\sqrt{2}}\frac{\tilde{m}_{\mu}}{\sqrt{1+x^2}} & 0  & 0 \\ \\
\frac{\tilde{m}_{e}(1+x^2)}{\sqrt{1+x^2-\tilde{m}_{\mu}^2}}e^{i\delta_{e}} & 0 & 0
\end{array}
\right)
\label{yu1}
\ee
and
\be
Y_{w}^{E2}=\frac{m_{\tau} }{v_2}\left( 
\begin{array}{ccc}
\frac{1}{\sqrt{2}}\frac{\tilde{m}_{\mu}}{\sqrt{1+x^2}} & 0 & 0 \\ \\
 0 &-\frac{1}{\sqrt{2}}\frac{\tilde{m}_{\mu}}{\sqrt{1+x^2}}  & \frac{1}{\sqrt{2}} \sqrt{\frac{1+x^2-\tilde{m}_{\mu}^2}{1+x^2}} \\ \\
0 & \frac{\tilde{m}_{e}(1+x^2)}{\sqrt{1+x^2-\tilde{m}_{\mu}^2}}e^{i\delta_{e}} & 0
\end{array}
\right).
\label{yu2}
\ee
In the computation of the flavour-changing neutral couplings, the
Yukawa couplings are defined in the mass basis which are obtained from
$Y_{w}^{EI}$, given in (\ref{yu1}) and (\ref{yu2}), according to
\[
\tilde{Y}_{m}^{EI}=U_{eL}^{\dagger}Y_{w}^{EI}U_{eR}
\]
where $U_{eL}$ and $U_{eR}$ are the matrices that diagonalize the
charged lepton mass matrix defined in eqs. (\ref{unu}) and
(\ref{unitary-leptons}).

We get,
\be
\tilde{Y}_{m}^{E1}\approx \frac{m_{\tau}}{v_{1}}\left(
\ba{ccc}
2\tilde{m}_{e} & -\frac{1}{2}\tilde{m}_{e} & \frac{1}{2} x \\
\\
-\tilde{m}_{\mu} & \frac{1}{2}\tilde{m}_{\mu} & -\frac{1}{2} \\
\\
\frac{1}{2} \tilde{m}_{\mu} x^2 & -\frac{1}{2}\tilde{m}_{\mu} & \frac{1}{2}
\ea
\right)_{m},
\label{y1m}
\ee
and
\be
\tilde{Y}_m^{E2}\approx \frac{m_{\tau}}{v_{2}}\left(
\ba{ccc}
-\tilde{m}_{e} & \frac{1}{2}\tilde{m}_{e} & -\frac{1}{2} x \\
\\
\tilde{m}_{\mu} & \frac{1}{2}\tilde{m}_{\mu} & \frac{1}{2} \\
\\
-\frac{1}{2} \tilde{m}_{\mu} x^2 & \frac{1}{2}\tilde{m}_{\mu} & \frac{1}{2}
\ea
\right)_{m},
\label{y2m}
\ee
where $\tilde{m}_{\mu}=5.94\times 10^{-2}$, $\tilde{m}_{e}=2.876 \times
10^{-4}$ and $x=m_{e}/m_{\mu}=4.84 \times 10^{-3}$.

All the nondiagonal elements in $\tilde{Y}_m^{EI}$ are responsible for
tree-level FCNC processes. The actual values of the Yukawa couplings
in eqs. (\ref{y1m}) and (\ref{y2m}) still depend on the VEV's of the
Higgs fields $v_{1}$ and $v_{2}$, and, hence, on the Higgs
potential. If the $S_{2}^{\prime}$ accidental symmetry in the Higgs sector is
preserved~\cite{pakvasa1}, $\langle H_{1}^{0}
\rangle = \langle H_{2}^{0} \rangle= v $. With the purpose of making
an order of magnitude estimate of the coefficient $m_\tau/v$
multiplying the Yukawa matrices, we may further assume that the VEV's
for all the Higgs fields are comparable, that is,
\[
\langle H_{s}^{0} \rangle=\langle
H_{1}^{0}\rangle = \langle H_{2}^{0}
\rangle=\frac{\sqrt{2}}{\sqrt{3}}\frac{M_W}{g_2}
\]
then
\[
m_\tau/v=\sqrt{3}/\sqrt{2}g_2m_\tau/M_W
\]
and we may estimate the numerical values of the Yukawa couplings from
the numerical values of the lepton masses.

Let us consider, first, the flavour violating process $\tau^-\to
\mu^-e^+e^-$, the amplitude of this process is proportional to $\tilde{Y}_{\tau \mu}^{EI}\tilde{Y}_{e e}^{EI}$~\cite{Sher:1991km}. Then,
the leptonic branching ratio for this process is

\be
Br(\tau \to \mu e^+ e^-)=\frac{\Gamma(\tau \to \mu e^+
  e^-)}{\Gamma(\tau \to e \nu \bar{\nu})+\Gamma(\tau \to \mu \nu \bar{\nu})}
\ee
where
\be
\Gamma(\tau \to \mu e^+
  e^-)\approx \frac{m_{\tau}^5}{3\times 2^{10} \pi^3}\frac{\left(\tilde{Y}_{\tau \mu}^{EI}\tilde{Y}_{e e}^{EI}\right)^2}{\left(M_{H_{I}}\right)^4}
\ee
which is the dominant term, the index I denotes the Higgs boson in the
$S_3$-doublet with the smallest mass. This equation, and the well-known
expression for $\Gamma(\tau \to e \nu \bar{\nu})$ and $\Gamma(\tau
\to \mu \nu \bar{\nu})$~\cite{PDG}, give
\be
Br(\tau \to \mu e^+
e^-)\approx\frac{9}{4}\left(\frac{m_{e}m_{\mu}}{m_{\tau}^2}\right)^2
\left(\frac{m_{\tau}}{M_{H_{1,2}}}\right)^4,
\ee
if we take for $M_{H_{1,2}}\sim 120~GeV$, we obtain
\be
Br(\tau
\to \mu e^+ e^-)\approx 3.15 \times 10^{-17},
\ee
well below the experimental upper bound for this process, which is $2.7 \times
10^{-8}$~\cite{Miyazaki:2007zw}. 
Similar computations give the following estimates
\be
Br(\tau \to e
\gamma)\approx
\frac{3\alpha}{8\pi}\left(\frac{m_\mu}{M_H}\right)^4,
\label{tauegamma}
\ee
\be
Br(\tau \to \mu
\gamma)\approx\frac{3\alpha}{128\pi}\left(\frac{m_{\mu}}{m_{\tau}}\right)^2\left(\frac{m_\tau}{M_H}\right)^4,
\ee
\be
Br(\tau \to 3\mu)\approx\frac{9}{64}\left(\frac{m_\mu}{M_H}\right)^4,
\ee
\be
Br(\mu \to 3e)\approx 18 \left(\frac{m_e
    m_\mu}{m_\tau^2}\right)^2\left(\frac{m_\tau}{M_H}\right)^4,
\label{mu3e}
\ee
and 
\be
Br(\mu \to e
\gamma)\approx
\frac{27\alpha}{64\pi}\left(\frac{m_e}{m_\mu}\right)^4\left(\frac{m_\tau}{M_H}\right)^4.
\label{muegamma}
\ee

From these equations, we see that FCNC processes in the leptonic sector are strongly
suppressed by the small values of the mass ratios
$m_e/m_\tau$, $m_\mu/m_\tau$ and
$m_\tau/M_H$. The numerical estimates of the branching
ratios and the corresponding experimental upper bounds are shown in
Table~\ref{table2}. It may be seen that, in all cases considered, the numerical
values for the branching ratios of the FCNC in the leptonic sector are
well below the corresponding experimental upper bounds.


\begin{table} \centering
\begin{tabular}{|l|l|l|l|}
\hline
FCNC processes & Theoretical BR &  Experimental  & References \\
& & upper bound BR &
\\ \hline
$\tau \to  3\mu$ & $8.43
\times 10^{-14}$& $ 3.2 \times 10^{-8}$ &  Y.~Miyazaki {\it et al.}
~\cite{Miyazaki:2007zw} 
\\ & &$5.3 \times 10^{-8}$ &  B. Aubert {\it et. al.} ~\cite{:2007pw}
\\ \hline
$\tau \to  \mu e^+ e^-$ & $3.15 \times 10^{-17}$& $2.7 \times 10^{-8} $ &Y.~Miyazaki {\it et al.} ~\cite{Miyazaki:2007zw}
\\ & & $8  \times 10^{-8}$ &  B. Aubert {\it et. al.} ~\cite{:2007pw}

\\ \hline

$\tau \to \mu \gamma$ &  $9.24 \times 10^{-15}$ & $ 6.8 \times
10^{-8}$& B. Aubert {\it et. al.} ~\cite{aubert2} 

\\ \hline
$\tau \to e \gamma$ & $5.22\times 10^{-16}$ & $ 1.1 \times 10^{-11}$ &  B. Aubert {\it et. al.} ~\cite{aubert3}  
\\ \hline 
$\mu \to  3e$ &  $2.53 \times 10^{-16}$ & $  1 \times 10^{-12}$ &
U. Bellgardt {\it et al.} ~\cite{bellgardt}  
\\ \hline
$\mu \to e \gamma$ &  $2.42 \times 10^{-20}$ & $ 1.2 \times 10^{-11}$& M.~L.~Brooks {\it et al.} ~\cite{Brooks:1999pu}
\\\hline
\end{tabular}
\caption{\label{table2}Leptonic FCNC processes, calculated with $M_{H_{1,2}}\sim 120~GeV$.}
\end{table}


\section{The anomalous magnetic moment of the muon}
\begin{figure}
\begin{center}
\includegraphics[width=4.5cm]{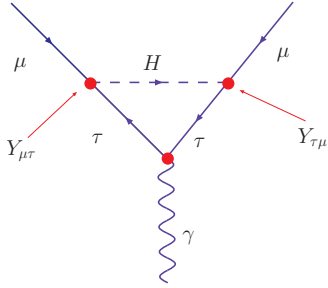}
\end{center}
\caption{\label{fig:anom}The contribution, $\delta a_{\mu}^{(H)}$, to the
  anomalous magnetic moment of the muon from the exchange of flavour
  changing scalars. The neutral Higgs boson can be a scalar or a pseudoscalar.}
\end{figure}

In models with more than one Higgs $SU(2)$ doublet, the exchange of
flavour changing scalars may contribute to the anomalous magnetic
moment of the muon. In the minimal $S_3$-invariant extension of the
Standard Model we are considering here, we have three Higgs
$SU(2)$ doublets, one in the singlet and the other two in the doublet
representations of the $S_3$ flavour group. The $Z_2$ symmetry
decouples the charged leptons from the Higgs boson in the $S_3$
singlet representation. Therefore, 
in the theory there are two neutral scalars and two neutral
pseudoscalars whose exchange will contribute to the anomalous magnetic
moment of the muon, in the leading order of magnitude. Since the
heavier generations have larger flavour-changing couplings, the
largest contribution comes from the heaviest charged leptons coupled
to the lightest of the neutral Higgs bosons, $\mu-\tau-H$, as shown in
Figure ~\ref{fig:anom}.
The contribution, $\delta a_{\mu}^{(H)}$, to the magnetic
moment of the muon from the exchange of the lightest neutral Higgs
boson computed in the leading order of magnitude is
\be
\delta a_\mu^{(H)}=\frac{\tilde{Y}_{\mu \tau} \tilde{Y}_{\tau \mu}}{16
  \pi^2}\frac{m_{\mu}m_{\tau}}{M_H^2}
\left(log\left(\frac{M_{H}^2}{m_{\tau}^2}\right)-\frac{3}{2}\right).
\ee
the Yukawa couplings appearing in this expression are given in
(\ref{y1m}) and (\ref{y2m}). Hence, we obtain
\be
\delta a_\mu^{(H)}=\frac{m_{\tau}^2}{(246~GeV)^2}\frac{(2+\tan^2\beta)}{32
  \pi^2}\frac{m_\mu^2}{M_{H}^2}\left(log\left(\frac{M_{H}^2}{m_{\tau}^2}\right)-\frac{3}{2}\right),
\label{g-2}
\ee
in this expression, $\tan \beta=  v_{s}/v_1$, is the ratio of the
vacuum expectation values of the Higgs scalars in the singlet
representation, $v_{s}$, and
in the doublet representation, $v_{1}$, of the $S_3$ flavour group. The most
restrictive upper bound on $\tan \beta$ may be obtained from the
experimental upper bound on $Br(\mu \to 3e)$ given in (\ref{mu3e}),
and in Table~\ref{table2}, we obtain
\be
\tan \beta \leq 14
\ee 
substitution of this value in (\ref{g-2}) and taking for the Higgs mass
the value $M_{H}=120~GeV$ gives an estimate of the
largest possible contribution of the FCNC to the anomaly of the
magnetic moment of the muon
\be
\delta a_{\mu}^{(H)}\approx 1.7 \times 10^{-10}.
\ee
This number has to be compared with the difference between the
experimental value and the Standard Model prediction for the anomaly
of the magnetic moment of the muon~\cite{Jegerlehner:2007xe}
\be
\Delta a_\mu= a_\mu^{exp}-a_\mu^{SM}=(28.7 \pm 9.1 )\times 10^{-10},
\ee
which means
\be
\frac{\delta a_{\mu}^{(H)}}{\Delta a_{\mu}}\approx 0.06.
\ee
Hence, the contribution of the flavour changing neutral currents to
the anomaly of the magnetic moment of the muon is smaller than or of the
order of $6\%$ of the discrepancy between the experimental value and
the Standard Model prediction. This discrepancy is of the order of
three standard deviations and quite important, but its interpretation
is compromised by uncertainties in the computation of higher order
hadronic effects mainly from three-loop vacuum polarization
effects, $a_{\mu}^{VP}(3,had)\approx-1.82\times 10^{-9}$~\cite{Erler:2006vu}, and
from three-loop contributions of hadronic light by light type,
$a_{\mu}^{LBL}(3,had)\approx 1.59\times 10^{-9}$~\cite{Erler:2006vu}. As explained above, the
contribution to the anomaly from flavour changing neutral currents in
the minimal $S_{3}$-invariant extension of the Standard Model,
computed in this work is, at most, $6\%$ of the discrepancy between
the experimental value and the Standard Model prediction for the anomaly, and is of the same order of magnitude as
the uncertainties in the higher order hadronic contributions, but
still it is not negligible and is certainly compatible with the best,
state of the art, experimental measurements and theoretical computations.
\section{Conclusions}

A variety of flavour violating effects related to the recent discovery
of neutrino oscillations and mixings was discussed in the framework of
the minimal $S_3$-invariant extension of the Standard Model that we
proposed recently~\cite{kubo1}. After a brief review of some relevant
results on lepton masses and mixings that had been derived in this
minimal $S_3$-invariant extension of the SM, we extended and updated
our previous results on the branching ratios of some selected
flavour-changing neutral currents processes
(FCNC)~\cite{Mondragon:2007af} as well as the contribution to the
magnetic moment of the muon~\cite{Mondragon:2007nk}. Some intersting
results are the following:
\begin{itemize}
\item The magnitudes of the three neutrino mixing angles,
  $\theta_{12}$, $\theta_{23}$ and $\theta_{13}$, are determined by an
  interplay of the $S_{3}\times Z_{2}$ symmetry, the see-saw mechanism and
the lepton mass hierarchy.
\item The neutrino mixing angles, $\theta_{23}$ and $\theta_{13}$,
  depend only on the masses of the charged leptons and their predicted
  numerical values are in excellent agreement with the best experimental values.
\item The solar mixing angle, $\theta_{12}$, fixes the scale and
  origin of the neutrino mass spectrum which has an inverted mass
  hierarchy with values
\[
\begin{array}{lcr}
m_{\nu_1}= 0.052~eV & m_{\nu_2}=0.053 ~eV & m_{\nu_3}= 0.019~eV.
\end{array}
\]
\item The branching ratios of all flavour changing neutral current processes
  in the leptonic sector are strongly suppressed by the $S_{3}\times Z_{2}$
  symmetry and powers of the small mass ratios $m_e/m_\tau$,
  $m_\mu/m_\tau$, and $\left(m_\tau/M_{H_{1,2}}\right)^4$, but could
  be important in astrophysical
  processes~\cite{Amanik:2004vm,Raffelt:2007nv}.
\footnote{There are in the literature extensions of the Standard Model
  with horizontal
  symmetries which offer explanations for inflation, baryogenesis and
  dark matter~\cite{Sakharov:1994pr}. An overview of
  the field of particle physics in astrophysics, prior to the
  discovery of neutrino oscillations and neutrino masses, where the
  above mentioned issues are discussed may be found in~\cite{Khlopov:1999rs}
}.

\item The anomalous magnetic moment of the muon gets a small but
  non-negligible contribution from the exchange of flavour changing
  scalar fields.
\end{itemize}

In conclusion, we may say that the minimal $S_3$-invariant extension
of the Standard Model describes successfully masses and mixings in the
quark~\cite{kubo1} (not discussed here) and leptonic sectors with a
small number of free parameters. It predicts the numerical values of
the $\theta_{23}$ and $\theta_{13}$ neutrino mixing angles, as well
as, all flavour changing neutral current processes in the leptonic
sector, in excellent agreement with experiment. In this model, the
exchange of flavour changing scalars gives a small but non-negligible
contribution to the anomaly of the magnetic moment of the muon.

\section{Acknowledgements}
We thank Prof. Jens Erler and Dr. Genaro Toledo-Sanchez for
helpful discussions about $g-2$. This work was partially supported by CONACYT M\'exico under contract
No 51554-F and by DGAPA-UNAM under contract PAPIIT-IN115207-2.


\end{document}